\def \SAIT #1 #2 {{\em Mem.\ Soc.\ Astron.\ It.\/} {\bf #1}, #2}
\def \MESS #1 #2 {{\em The Messenger\/} {\bf #1}, #2}
\def \ASTRNACH #1 #2 {{\em Astron. Nach.\/} {\bf #1}, #2}
\def \AAP #1 #2 {{\em Astron. Astrophys.\/} {\bf #1}, #2}
\def \AAL #1 #2 {{\em Astron. Astrophys. Lett.\/} {\bf #1}, L#2}
\def \AAR #1 #2 {{\em Astron. Astrophys. Rev.\/} {\bf #1}, #2}
\def \AAS #1 #2 {{\em Astron. Astrophys. Suppl. Ser.\/} {\bf #1}, #2}
\def \AJ #1 #2 {{\em Astron. J.\/} {\bf #1}, #2}
\def \ANNREV #1 #2 {{\em Ann. Rev. Astron. Astrophys.\/} {\bf #1}, #2}
\def \APJ #1 #2 {{\em Astrophys. J.\/} {\bf #1}, #2}
\def \APJL #1 #2 {{\em Astrophys. J. Lett.\/} {\bf #1}, L#2}
\def \APJS #1 #2 {{\em Astrophys. J. Suppl.\/} {\bf #1}, #2}
\def \APSS #1 #2 {{\em Astrophys. Space Sci.\/} {\bf #1}, #2}
\def \ASR #1 #2 {{\em Adv. Space Res.\/} {\bf #1}, #2}
\def \BAIC #1 #2 {{\em Bull. Astron. Inst. Czechosl.\/} {\bf #1}, #2}
\def \JSQRT #1 #2 {{\em J. Quant. Spectrosc. Radiat. Transfer\/} {\bf #1}, #2}
\def \MN #1 #2 {{\em Mon. Not. R. Astr. Soc.\/} {\bf #1}, #2}
\def \MEM #1 #2 {{\em Mem. R. Astr. Soc.\/} {\bf #1}, #2}
\def \PLR #1 #2 {{\em Phys. Lett. Rev.\/} {\bf #1}, #2}
\def \PASJ #1 #2 {{\em Publ. Astron. Soc. Japan\/} {\bf #1}, #2}
\def \PASP #1 #2 {{\em Publ. Astr. Soc. Pacific\/} {\bf #1}, #2}
\def \NAT #1 #2 {{\em Nature\/} {\bf #1}, #2}

\documentstyle{memsait}
\begin{opening}
\title{{\it RXTE} OBSERVATIONS OF PSR J1105$-$6107} 
\author{JULIA STEINBERGER$^1$, VICTORIA M. KASPI$^1$, ERIC V. GOTTHELF$^2$}
\institute{$^1$Department of Physics and Center
for Space Research, Massachusetts Institute of Technology, Cambridge, MA, USA 02139\\
$^2$Goddard Space Flight Center, Greenbelt, MD, USA 20771}
\date{} 
\end{opening}

\begin{document}

\oddpagefooter{}{}{} 
\evenpagefooter{}{}{} 
\ 

\begin{abstract}
We report on X-ray observations of the young, energetic radio pulsar
PSR J1105$-$6107 obtained using the {\it Rossi X-ray Timing Explorer}.
Our timing analysis of Proportional Counter Array
data finds no evidence for X-ray pulsations from this source.  We set a
new upper limit on the pulsed X-ray flux from the pulsar of
18\% in the 2-10~keV band, for a duty cycle of 0.5.
Our improved upper limit 
supports the hypothesis that the X-ray emission seen by the {\it ASCA}
X-ray satellite in the direction of PSR~J1105$-$6107 is the result of a
pulsar-powered synchrotron nebula.
\end{abstract}

\section{Introduction}

The radio pulsar PSR J1105$-$6107 has a spin period of $P=63$~ms and a
characteristic spin-down age $P/2\dot{P}= 63$ kyr (Kaspi
et al. 1997).  The pulsar's spin-down luminosity $\dot{E} \equiv 4 \pi^2 I
\dot{P} / P^3 = 2.5\times10^{36} {\rm erg}\; {\rm s}^{-1}$, where $I$
is the neutron star moment of inertia, assumed to be
$10^{45}$~g~cm$^2$.  The pulsar's distance from the Earth is 7 kpc,
estimated from its dispersion measure (Taylor \& Cordes 1993).  Given
its $\dot{E}$ and distance, PSR~J1105$-$6107 is expected to be an
observable source of high energy emission.  Indeed, the pulsar has been
detected by the {\it ASCA} X-ray satellite as a faint, unpulsed point
source (Gotthelf \& Kaspi 1998, hereafter GK98).   The pulsed fraction
upper limit set using those data is 31\% for a duty cycle of 0.5 in the
2-10~keV band.

GK98 concluded that the X-rays observed by {\it ASCA} are due to a pulsar-powered
synchrotron nebula, the result of confinement of the pulsar wind by
the ram pressure of the surrounding interstellar medium.
Other possible X-ray emission mechanisms from the pulsar
include thermal emission from the neutron star surface and pulsed
non-thermal emission from the magnetosphere.  Only the latter
could contribute to the {\it ASCA} source; determining at what level
is important for understanding
this mechanism, as well as for observationally establishing the nebular
flux and spectrum and ultimately the overall energy budget
of the pulsar and the physics of the nebula,
as well as for planning strategies for follow-up
observations.

Here we report on a long observation in the direction of PSR
J1105$-$6107 done by the {\it Rossi X-Ray Timing Explorer} ({\it
RXTE}).  Our goal was to detect, or place upper limits on, pulsed
X-ray emission in order to better constrain the nature of the 
{\it ASCA}-detected X-ray source.

\section{Observations and Timing Analysis}

PSR J1105$-$6107 was observed by {\it RXTE} on  November 17--19, 1997 
(MJD 50769--50771).  We report on observations made with the Proportional 
Counter Array (PCA).
The total observing time was 100 ks.\footnote{During most of the
observation, all 5 Proportional Counter Units were on, however, PCUs 4
and 5 were turned off for 1~hr each for detector management.} Each of
the 5 Proportional Counter Units (PCUs) has a maximum effective area of
$\sim$1275~cm$^2$, peaking around 10 keV. The total energy
range of the PCA is 2$-$100 keV.  
The PCA has no spatial resolution and a large field of view of
1 square degree.   For more detailed information regarding the PCA and
{\it RXTE} see {\tt http://heasarc.gsfc.nasa.gov/docs/frams/xte\_geninfo.html.}

The data were taken in Standard 1 and 2, and Good Xenon Experiment Data
System (EDS) modes (see Jahoda et al. 1996). The Good Xenon data have a
time resolution of 1~$\mu$s and are used for the timing
analysis.  The Standard 2 data could be used for a spectral analysis, however
one or more bright sources contaminate the field of view and preclude any spectral
analysis useful for constraining the pulsar emission.

The PCA Good Xenon packet data were  merged, and binned into a single
time series of resolution 2~ms.  Energy channels were then combined,
cleaned of Earth occults and event times reduced to the barycenter.
The time series was then folded at the observation-contemporary radio
ephemeris obtained at the Parkes 64-m radio telescope in New South
Wales, Australia.  The radio ephemeris used is given in Table~1 and
was determined using observations similar to those described
by Kaspi et al. (1997).  In addition, a timing analysis was carried out
using the {\it RXTE} {\tt ftool fasebin} (e.g.  Rots et al. 1998),
which folds the (barycentered, clock corrected) Good Xenon data
using the pulsar ephemeris.
In both cases, only the first detector layer events were used to maximize
signal-to-noise ratio as our source is not bright.

\vspace{1cm} 
\centerline{\bf Tab. 1 - Radio Pulse Ephemeris for PSR J1105$-$6107}
 
\begin{table}[h]
\hspace{3.0cm} 
\begin{tabular}{|c|c|}
\hline
Frequency & 15.8245806130610 Hz \\
Frequency Derivative & $-$3.96320$\times 10^{-12}$ Hz s$^{-1}$\\
Epoch & MJD 50711 \\
Range of Validity & MJD 50638--50785 \\
\hline
\end{tabular}
\end{table}

No X-ray pulsations were found.  The time series was also folded using
data from several restricted energy ranges: the complete {\it RXTE} range
(2--100 keV), the {\it ASCA} band (2--10 keV), and the {\it RXTE} maximum effective
area range (5--19 keV), as well is in several selected narrower ranges.  
In all cases, the $\chi^2$ of the folded
profile indicated that it was consistent with being flat.

Although the radio ephemeris gives an accurate pulsar period for the
time of the observation, we also performed period searches around the
known pulsar period.  The total time span of the observation is $T =
134177$ s. The largest adequate frequency interval for searching is
$\Delta \nu = (T \times N_{b} )^{-1}$ where $N_{b}$ is the number of
phase bins used.  For $N_{b}=20$, $\Delta \nu=3.7\times10^{-7}$~Hz.
We folded the data at 550 frequencies around the known pulsar radio
frequency, from 15.8245 to 15.8247~Hz.  This analysis also
revealed no significant signals.

Here we derive a pulsed flux upper limit for our {\it RXTE}
observation.  Using the binning-independent H test, optimal for use
when the pulse profile is unknown (de Jager 1994),
we obtain a 3$\sigma$ upper limit on the
pulsed fraction of all detected counts.
We assume the emission can be described by a
power-law spectrum with photon index 2, consistent with
the results of GK98.  We parametrize the PCA effective area by
an approximate analytic form in the 2--20~keV band, $A(E) \simeq 5700
\exp(-(E-7.4)^2/32)$~cm$^2$, with $E$ in keV.
The 3$\sigma$ upper limit on the pulsed flux is
\begin{equation}
{\rm f}_{ul}(3\sigma) = 1.602 \times 10^{-9}
\frac{N_0}{t}
\int^{E_{max}}_{E_{min}}\frac{E}{A(E)}E^{-2}dE \;\; {\rm erg \; cm^{-2} \; s^{-1}},
\end{equation}
where $E$ is in keV.
$N_0$ is a normalization factor for the spectrum determined from
the pulsed fraction upper limit and the total number
of counts detected, and $t$ is the total good integration time, 98.5~ks.
This analysis does not take into account the actual spectrum of
the {\it RXTE} signal (including background).
In the {\it ASCA} band of 2--10 keV, assuming a pessimistic 0.5 duty cycle,
we find a 3$\sigma$ upper limit on the pulsed flux of 
$1.15\times10^{-13}$~erg~cm$^{-2}$~s$^{-1}$.
This upper limit corresponds to
$18\%$ of the flux observed by {\it ASCA} in the 2--10~keV
band.  By contrast, the upper limit on the pulsed flux from the {\it ASCA}
observation is $31\%$ in the same energy range, for the same duty cycle.
For duty cycle 0.3, our revised upper limit is 13\% in the {\it ASCA} band.


\section{Summary and Discussion} 

We have obtained an improved X-ray pulsed flux upper limit of 18\% in
the 2--10~keV band for duty cycle 0.5 from the young, energetic radio
pulsar PSR~J1105$-$6107.  This represents a significant decrease in the
pulsed fraction upper limit derived by GK98 using {\it ASCA} data.  The
absence of X-ray pulsations in our {\it RXTE} observation of
PSR~J1105$-$6107 is consistent with results seen for other pulsars of
its type, namely those that have characteristic ages greater than 
$\sim$10~kyr.  This group of pulsars has been dubbed ``Vela-like,'' because
they exhibit an absence of sharp, non-thermal magnetospheric
pulsations as are seen in the much younger sources like the Crab
pulsar, and because they show prominent X-ray synchrotron emission (see
Becker \& Tr\"umper 1997 for a discussion).  Thus, the X-ray emission
from PSR~J1105$-$6107 likely has its origin in a pulsar-powered
synchrotron nebula.  GK98 suggested the nebula may formed by the
ram-pressure confinement of the relativistic pulsar wind by a high
pulsar space velocity.  This hypothesis can be verified by resolving
the morphology of the putative nebula, using the next generation of
X-ray telescopes, in particular {\it AXAF} and {\it XMM}.

\acknowledgements
We thank K. Jahoda as well as the {\it RXTE} team at MIT and for help
and resources, particularly D. Chakrabarty, E. Morgan, and M. Muno.
Thanks also to D.  Manchester, R. Pace, and M. Bailes for supporting
radio observations at the Parkes observatory.  The research was
supported by NASA grant NAG5-7131.


\end{document}